\documentclass[12pt,a4paper]{article}

\usepackage{amsmath}
\usepackage{caption}
\usepackage{subcaption}
\usepackage{indent first}
\usepackage{jheppub}
\usepackage{tikz}

\newcommand{\bC}{\mathbb{C}}
\newcommand{\bR}{\mathbb{R}}
\newcommand{\bZ}{\mathbb{Z}}
\def\Nequals#1{$\mathcal{N}{=}#1$}
\def\SU{\mathrm{SU}}
\def\U{\mathrm{U}}
\def\SO{\mathrm{SO}}
\def\tr{\mathop{\mathrm{tr}}\nolimits}
\def\Tr{\mathop{\mathrm{Tr}}\nolimits}

\usepackage{xcolor}

\begin{document}

\title{Anomaly polynomial of E-string theories}
\abstract{We determine the anomaly polynomial of 
the E-string theory and its higher-rank generalizations, that is, the 6d \Nequals{(1,0)} superconformal theories
on the  worldvolume of one or multiple M5-branes embedded within the end-of-the-world brane with $E_8$ symmetry.
}
\author[1]{Kantaro Ohmori,}
\author[1]{Hiroyuki Shimizu,}
\author[1,2]{and Yuji Tachikawa}
\affiliation[1]{Department of Physics, Faculty of Science, \\
 University of Tokyo,  Bunkyo-ku, Tokyo 133-0022, Japan}
\affiliation[2]{Institute for the Physics and Mathematics of the Universe, \\
 University of Tokyo,  Kashiwa, Chiba 277-8583, Japan}
\preprint{IPMU-14-0097, UT-14-18}

\maketitle

\section{Introduction}\label{Introduction}
In the last few years, there has been a significant progress in our understanding of 6d \Nequals{(2,0)} superconformal theories and their compactifications to lower dimensions, starting with \cite{Gaiotto:2009we}.   
The dynamics of 6d \Nequals{(1,0)} superconformal theories, however, remains quite mysterious. 

One class of 6d \Nequals{(1,0)} theories is obtained by taking the decoupling limit of $Q$ coincident  M5-branes  embedded within the $E_8$ end-of-the-world brane of M-theory.  When $Q=1$, the theory is commonly known as \emph{the E-string theory}, as the stringy degrees of freedom in this theory has $E_8$ flavor symmetry.\footnote{  
The properties of this theory in 6d were studied e.g.~in \cite{Ganor:1996mu,Seiberg:1996vs,Witten:1996qb,Morrison:1996pp,Witten:1996qz,Cheung:1997id}. The dynamics of this theory on $S^1$ or $T^2$ was rather extensively studied, 
but we do not cite them here. 
}
We call the theories for $Q{>}1$ \emph{the E-string theories of general rank}. 
The objective of this paper is to compute their anomaly polynomials, thereby adding an item to the list of known properties of these mysterious theories. 

Let us quickly recall the symmetry of E-string theories. As we already mentioned, they have \Nequals{(1,0)} superconformal symmetry and $E_8$ flavor symmetry. The transverse space to $Q$ M5 branes has the form $\bR^4\times \bR_{>0}$, and therefore they have $\SO(4)\simeq \SU(2)_L\times \SU(2)_R$ symmetry. One of the two $\SU(2)$ symmetries is the R-symmetry in the superconformal algebra.
 The low-energy limit of this brane system consists of a decoupled single free hypermultiplet, describing the center-of-mass motion of $Q$ M5 branes within the $E_8$ end-of-the-world brane, and a genuinely interacting 6d superconformal field theory.

Without further ado, here we will present the final result.
The anomaly polynomial of the total system, including the contribution from the free hypermultiplet is
\begin{equation}
	A_{E_8+\text{free}}(Q)=Q^3 \frac{p_2(N)}6
	+ Q^2 \frac{\chi_4(N) I_4}{2} +
	Q  \left(\frac{I_4^2}{2}-I_8\right)
	\label{main}
\end{equation} 
where 
\begin{align}
	I_4&=\frac{1}{4}\left(p_1(N)+p_1(T) + \Tr  F^2\right),\\
	I_8&=\frac{1}{48}\left(p_2(N)+p_2(T) -\frac14(p_1(N)-p_1(T))^2\right).
\end{align}

Here we used the symbols $F$ for the $E_8$ background field, $T$ for the tangent bundle of the worldvolume and  $N$ for the $\SO(4)$ normal bundle; $p_i$ are the Pontrjagin classes and $\chi_4(N)$ is the Euler class.\footnote{Our normalization of the anomaly polynomials is such that the contribution of a Weyl fermion in a gauge representation $\rho$ is $\hat A(T)\tr_\rho e^{iF}$. In particular, we take $F$ to be anti-hermitian and we include a factor $(2\pi)^{-1}$ in the definition of $F$.}
Our $\Tr$ is the trace in the adjoint representation divided by the dual Coxeter number. Therefore, the integral of $\Tr F^2/4$ over a four-cycle gives the instanton number in the standard normalization. 

Under the decomposition $\SO(4)\simeq \SU(2)_R \times \SU(2)_L$, we have\footnote{In our convention, a positively charged M5-brane has instanton number 1, and preserves the same supersymmetry as the K3 manifold in the standard orientation. As $\int_\text{K3} p_1=-48$ and $\int_\text{K3}\chi_4=24$, we have $c_2(L)=24$ and $c_2(R)=0$. This means that $c_2(R)$ corresponds to the R-symmetry of the 6d supersymmetry. } \begin{equation}
p_1(N)=-2(c_2(R)+c_2(L)),\quad
\chi_4(N)=c_2(L)-c_2(R),\quad 
p_2(N)=\chi_4(N)^2
\end{equation} where $c_2(L)$, $c_2(R)$ are the second Chern classes of the rank-2 bundles $L$, $R$ such that $L\otimes R \simeq N_\bC$. 
When $Q>0$, $\SU(2)_R$ is the R-symmetry and $\SU(2)_L$ is a flavor symmetry; when $Q<0$ the assignment is reversed. In the following we assume $Q>0$ unless otherwise specified. The anomaly polynomial of the system without the decoupled center-of-mass part is obtained by subtracting from \eqref{main} the contribution of the free hypermultiplet, which is a half-hypermultiplet in the doublet of $\SU(2)_L$. This is given by  \begin{equation}
A_\text{free}=\frac{7p_1(T)^2-4p_2(T)}{5760} +\frac{c_2(L) p_1(T)}{48} + \frac{c_2(L)^2}{24}.\label{freehyper}
\end{equation}

The rest of the paper is organized as follows. 
In Sec.~\ref{computation}, we will compute  the anomaly polynomial  by combining the analysis of Ho\v rava and Witten \cite{Horava:1996ma} of the anomaly of the $E_8$ end-of-the-world brane and that of Freed, Harvey, Minasian and Moore \cite{Freed:1998tg,Harvey:1998bx} of the anomaly of multiple coincident five-branes. 
In Sec.~\ref{checks}, we perform three checks of the computation. 
First, we give another derivation of the terms in \eqref{main} that do not involve the normal bundle using the compactification of heterotic string theory on K3. Second, we compare the coefficient of $(\Tr F^2)^2$ with that computed in \cite{Witten:1996qz,Cheung:1997id}. Third, we show that when $Q=1$ the $c_2(L)$ dependence of \eqref{main} comes solely from the free hypermultiplet.
We conclude the paper in Sec.~\ref{conclusions} by discussing how the anomaly polynomial might be used.

In this paper,  we will compute only the part of the anomaly that can be captured at the level of de Rham cohomology. For this purpose, the methods of \cite{Horava:1996ma,Freed:1998tg,Harvey:1998bx}  suffice. To obtain the information on the global anomaly, we might need more sophisticated methods that can be found  e.g.~in \cite{Diaconescu:2000wy,Diaconescu:2003bm}. We note that the global anomaly of  M5-branes was analyzed in \cite{Monnier:2013rpa}.
 
\section{Computations}\label{computation}

\subsection{Chern-Simons terms of M-theory}

The M-theory contains two kinds of Chern-Simons terms. 
The first term is the Chern-Simons coupling of the eleven-dimensional supergravity\footnote{Our sign convention of the Chern-Simons terms is the negative of that in \cite{Witten:1996md}. With this choice, the overall sign of the anomaly polynomials of $Q$ M5-branes reproduces the one reported in \cite{Harvey:1998bx}. For a through discussion of issues of conventions in M-theory Chern-Simons couplings, see \cite{Bilal:2003es}. 
 }:
\begin{equation}
S_{CGG}=\frac{2\pi}{6}\int_{X_{11}}  C \wedge  G  \wedge  G 
\label{eq:CGGcoupling}
\end{equation}
where $X_{11}$ is the 11d manifold on which the M-theory is defined, $ C $ is the 3-form potential and $ G $ is the 4-form field strength of $ C $. We normalize $ G $  so that $\int_S G \in \bZ$ for four-cycles $S$ with  $\int_S w_4=0$. 

We prefer to represent this coupling  $\eqref{eq:CGGcoupling}$ as 
\begin{align}
	S_{CGG}=\frac{2\pi}{6}\int_{Y_{12}} G \wedge G \wedge  G .
	\label{eq:GGGcoupling}
\end{align}
where $Y_{12}$ is a 12d manifold whose boundary $\partial Y_{12}$
is equal to $X_{11}$.
  
The second term is 
  \begin{equation}
  S_{CI_{8}} =- 2\pi \int_{X_{11}} C \wedge I_{8}, \qquad
  I_{8}=\frac{1}{48} \biggl{[}p_{2}(TX_{11})-\frac{1}{4}p^{2}_{1}(TX_{11}) \biggr{]}
  \end{equation}
  where  $p_{i}$ denotes the $i$-th Pontrjagin class.
Again we rewrite this term using $Y_{12}$:
\begin{align}
	S_{CI_8} =- 2\pi \int_{Y_{12}}  G \wedge  I_8.
	\label{eq:GI8coupling}
\end{align}
  
The existence of this interaction $S_{CI_{8}}$, which  is not present in the naive supergravity action up to two derivatives, is known from various points of view, including (but not limited to) the following:
First, by the reduction to the type IIA, this interaction produces $B \wedge I_{8}$ coupling,
which is known to be generated by a one-loop effect in type IIA superstring theory \cite{Vafa:1995fj}. 
Second, this interaction is necessary for the gravitational anomaly cancellation of a single M5-brane \cite{Duff:1995wd,Witten:1995em}. Third, on general manifolds, $S_{CGG}$ is  well-defined only when accompanied with $S_{CI_8}$ \cite{Witten:1996md}.
 
With these Chern-Simons couplings, we can calculate the anomalies due to degrees of freedom on  M5-branes or on $E_8$ end-of-the-world branes, or combinations thereof.  
In the next two subsections we review the computations of the anomaly polynomials of $Q$ coincident M5-branes and of one $E_8$ end-of-the-world brane, respectively.  An experienced reader can go directly to Sec.~\ref{sec:EstringAnom}.
  
\subsection{Anomalies of M5-branes}
Let us first review the procedure of \cite{Freed:1998tg,Harvey:1998bx} and obtain the anomaly polynomial for $Q$ coincident M5-branes. 

Consider the M-theory with $Q$ M5-branes on $X_6$.
We take a coordinate $x_i, i=1,\cdots,11$ such that $X_6=\{x_7=x_8=x_9=x_{10}=x_{11}=0\}$.
The Bianchi identity for 4-form field strength $ G $ becomes
\begin{align}
	 d  G =Q\prod_{i=7}^{11}\delta(x_i) d x_i,
	\label{eq:Bianchidelta}
\end{align}
because M5-branes are magnetic source for $ G $.

In the presence M5-branes, the Lagrangian density of the bulk Chern-Simons terms \eqref{eq:GGGcoupling} and \eqref{eq:GI8coupling} becomes singular around the worldvolume $X_6$.
Such contribution gives rise to an anomaly inflow toward $X_6$, 
which should be cancelled by the anomalies carried by the degrees of freedom on M5-branes, allowing us to determine the anomalies. 

To carry out this computation, we first  regularize the singularity appropriately.  We follow \cite{Freed:1998tg} and modify the Bianchi identity to be 
\begin{equation}
	 d  G = Q  d \rho\: e_4/2.
	\label{eq:modifiedBianchiFHMM}
\end{equation}
Here, $e_4$ is the global angular form of the normal bundle of $X_6$ with the normalization $\int_{S^4} e_4=2$, and $\rho=\rho(r)$ is a bump function which depends on the distance $r$ from $X_6$ and satisfies $\rho(0)=-1$ and $\rho(r) =0$ when $r$ is sufficiently large. 
The properties of the global angular form are summarized in Appendix.~\ref{appendix}.
With this modification, we have a following regular solution for \eqref{eq:modifiedBianchiFHMM}:
\begin{align}
	 G = d  C - Q  d \rho\:e_3^{(0)}/2.
\end{align}

The Chern-Simons couplings \eqref{eq:GGGcoupling} and \eqref{eq:GI8coupling} also require a modification, as $G$ is no longer closed. 
The proposal in \cite{Freed:1998tg} in our notation is the following.
Let $Y_7$ be a submanifold of $Y_{12}$ whose boundary is $X_6$. 
Let $G'$ be a closed version of $G$ given by  
\begin{align}
	  G '= G -Q \rho e_4/2
\end{align} defined throughout $Y_{12}$. To avoid the singularity of $ G '$ at $Y_7$,
we refine the integration procedure as: \begin{align}
	\int_{Y_{12}}\to 
	\lim_{\epsilon\to 0}\int_{Y_{12}\setminus  D _{\epsilon}(Y_7)}.
\end{align}
Here and in the following,  $D_{\epsilon}(M)$ for a space $M$ denotes the tubular neighborhood of $M$  with radius $\epsilon$ in general. The orientation is such that $\partial(Y_{12}\setminus D_\epsilon(Y_7))=-\partial D_\epsilon(Y_7)$.

The proper Chern-Simons couplings are then finally given by 
\begin{align}
	S_{CGG}=\frac{2\pi}{6}
	\lim_{\epsilon\to 0}\int_{Y_{12}\setminus  D _{\epsilon}(Y_7)}
	 G '\wedge  G '\wedge  G ',
	\label{eq:GGGcoupling2}
\end{align}
and 
\begin{align}
	S_{CI_8}= -2\pi
	\lim_{\epsilon\to 0}\int_{Y_{12}\setminus  D _{\epsilon}(Y_7)}
	 G '\wedge I_8.
\end{align}

To calculate the anomalies, we concentrate on the most singular part of these terms.
For $S_{CGG}$, the singular part (which is independent of $ C $) becomes 
\begin{align}
	S_{CGG}|_{\text{sing}}
	&=
	-\frac{2\pi Q}{6} \lim_{\epsilon\to 0}\int_{\partial{D}_\epsilon(Y_7)}
	\biggl{(}-(\rho e_3^{(0)})/2 \wedge  G '|_{\text{sing}} \wedge  G '|_{\text{sing}} \biggr{)}\nonumber\\
	&=
	-\frac{2\pi Q^3}{6\cdot 8}\lim_{\epsilon\to 0}\int_{\partial{D}_\epsilon(Y_7)} (-\rho(\epsilon))^3 e_3^{(0)}e_4^2  
	= -\frac{2\pi Q^3}{24}\int_{Y_7} p_2^{(0)}(N)
\end{align}
where $N$ denotes the normal bundle and we used the formula \eqref{eq:BottCattaneo1} in the last line.

Similarly, the singular part of $S_{CI_8}$ gives
\begin{align}
	S_{CI_8}|_{\text{sing}}&=2\pi Q
	\lim_{\epsilon\to 0}\int_{\partial{D}_\epsilon(Y_7)}
	\biggl{(} -(\rho e_3^{(0)})/2 \wedge I_8 \biggr{)}
	 = 2\pi Q \int_{Y_7} I_7^{(0)}.
\end{align}

Therefore, the contribution to the anomaly polynomial from the bulk Chern-Simons terms is 
\begin{align}
	- \frac{Q^3}{24} p_2(N)+ Q I_8,
\end{align} 
and the anomaly polynomial $A_{\text{M5}}(Q)$ of the field theory on $Q$ coincident M5-branes is its negative, that is,
\begin{align}
	A_{\text{M5}}(Q)=\frac{Q^3}{24} p_2(N) - Q I_8.
\end{align}

\subsection{Anomalies of the $E_8$ end-of-the-world brane} \label{subsec:M9anomaly}
Let us use the formalism reviewed so far to reproduce the computation of Ho\v rava and Witten \cite{Horava:1996ma}.
To introduce an end-of-the-world  brane, we take a $\bZ_2$ orbifold $X_{11}/\bZ_2$ whose fixed points contain a ten-dimensional component $X_{10}$.
For definiteness, we assume the $\bZ_2$ action sends $x_{11}$ to $-x_{11}$.

In the presence of the end-of-the-world brane, the Bianchi identity is modified as
 \begin{equation}
	 d G =  d \sigma e_{0}I_{4}.
 \label{eq:modifiedBianchiHW}
 \end{equation}
In this expression, we smoothed out the delta function $\delta(x_{11}) d x_{11}$ to the Thom class $ d (\sigma e_{0}/2)$.  Note that $e_0$ is just a step function which is $1$ for $x_{11}>0$ and $-1$ for $x_{11}<0$, and $\sigma$ is a bump function which is $-1$ at $x_{11}=0$ and $0$ sufficiently far away.  Finally, $I_4$ is  \begin{align}
	 I_4=\frac{1}{4}(\Tr F^2 + p_1(TX_{11}))
 \end{align}
where $F$ is the field strength of $E_8$ gauge field on the end-of-the-world brane. 

We extend the orbifold action to the auxiliary space $Y_{12}$, and denote by  $Y_{11}$ a component of fixed points whose boundary is $X_{10}$. 
Repeating  the same calculation as in the previous subsection using \begin{equation}
G=dC-d\sigma e_0 I_3^{(0)},\qquad G'=G-\sigma e_0 I_4,
\end{equation}
we get the singular part of the Chern-Simons terms to be \begin{equation}
S_{CGG}|_{\text{sing}} + S_{CI_8}|_{\text{sing}} 
= -\int_{Y_{11}} I_3^{(0)} (\frac{1}{6} I_{4}^{2}-I_{8}).
\end{equation} 
Therefore the anomaly polynomial on the end-of-the-world brane is
\begin{equation}
	A_{\text{$E_8$-brane}}= I_{4}(\frac{1}{6} I_{4}^{2}-I_{8}).
	\label{eq:M9anomaly}
\end{equation}
This reproduces the anomaly of an $E_8$ vector multiplet plus one half the anomaly of the supergravity multiplet, as discussed in \cite{Horava:1996ma}. For example, the pure gauge term of the anomaly of the $E_8$ vector multiplet is $(1/2)\tr_\text{adj} F^6/6!$,
which equals $(\Tr F^2/4)^3/6$
using the identity $\tr_\text{adj} F^6 = (15/4) (\Tr F^2)^3$. 

\subsection{E-string anomalies}\label{sec:EstringAnom}

Let us now compute  the anomaly polynomial of the E-string theory of rank $Q$,
which is the field theory on $Q$ coincident M5-branes in the $E_8$ end-of-the-world brane.
In addition to the set-up in the last subsection, we put $Q$ M5-branes onto $X_6 \subset X_{10}$ .

 We modify the Bianchi identity for $ G $ to be
\begin{equation}
	 d  G =  d \sigma  e_{0}I_4 + 2Q  d \rho  \frac{e_4}2,
	\label{eq:modifiedBianchiEst}
\end{equation}
combining \eqref{eq:modifiedBianchiFHMM} and \eqref{eq:modifiedBianchiHW}.
There are two bump functions $\sigma $ and $\rho $: $\sigma$ is for the end-of-the-world brane and is   $\bZ_2$-symmetric, and $\rho$ is for M5-branes and  $\SO(5)$ covariant. 
Note that the flux due to M5-branes should be $2Q$, as we need to include the one  from mirror images.

We note that our Bianchi identity \eqref{eq:modifiedBianchiEst} naturally incorporates the property that
one M5-brane can dissolve into the end-of-the-world brane 
to become a small 1-instanton configuration of the $E_8$ gauge configuration.
Indeed, consider an $E_8$ gauge field having $k$ zero-sized instantons along $X_6$. Then, $ d \sigma  I_4 e_0$ behaves as $2k  d \rho  e_{4}/2$,  in that both behave as $2k$ times a Poincar\'e dual of $X_6$.   
In \eqref{eq:modifiedBianchiEst}, this is equivalent to change the number of M5-branes by $Q \rightarrow Q + k$, as expected.
In the following, we put the contribution from zero-sized instantons into $Q$,
and assume  $I_4$ is regular on $X_{10}$.

We solve the Bianchi identity \eqref{eq:modifiedBianchiEst} so that the field strength $ G $ has no singularity:
\begin{equation}
	 G  =  d  C  - d \sigma  e_{0} {I}_{3}^{(0)}  - Q d  \rho {e}_{3}^{(0)}.
\end{equation}
We now need a closed version $G'$ of $G$, which is
\begin{equation}
	 G ' =  d  C  - d (\sigma  {I}^{(0)}_3 e_0)  - Q d  (\rho e_{3}^{(0)}) .
	\label{eq:modifiedG}
\end{equation}
This modified field strength has two types of singularities:
the second term is singular along $Y_{11}$, and the third term is singular along $Y_{7}$.

Then, the properly modified Chern-Simons terms in the action are:
\begin{equation}
	S_{CS}=\lim_{\epsilon_{1,2}\to 0}2\pi \int_{Y_{12}\setminus ({D}_{\epsilon_1}(Y_{11})\cup {D}_{\epsilon_2}(Y_7))/\bZ_2}
	\left(\frac{1}{6} G '\wedge  G '\wedge  G ' -   G ' \wedge I_{8}\right).
	\label{eq:modifiedSCS}
\end{equation}
The singularities which do not contain the third term of \eqref{eq:modifiedG} is the same as what we  calculated in subsection \ref{subsec:M9anomaly}, and are cancelled by the anomalies on the $E_8$ end-of-the-world brane.
The remaining singular part of \eqref{eq:modifiedSCS} localizes to the boundary of $D_{\epsilon_2}(Y_7)$, and should cancel for the anomalies of the E-string theory of rank $Q$. This remainder is given by 
\begin{align}
	&S_{CS}|_{\text{sing}}+\int_{Y_{11}} A_{\text{$E_8$-brane}}^{{(0)}}\nonumber\\
		&= 2\pi \int_{\partial D(Y_7)/\bZ_2 }
	\left(-\frac{Q^3}{6}  e_3^{(0)} e_4^2 -  \frac{Q^2}{2} e_3^{(0)} e_4  I_4 e_0- \frac{Q}{2} e_3^{(0)} I_4^2
	+ Q e_3^{(0)} I_8\right)\nonumber\\
	&=2\pi\int_{Y_7} \left( -\frac{Q^3}{6} p_2^{(0)} -\frac{Q^2}{2}\chi_4 I_3^{(0)} -\frac{Q}{2} I_4 I_3^{(0)}
	+ Q I_7^{(0)}\right).\label{intermediate}
\end{align}
In the last line, we used the integration formulas of the global angular forms \eqref{eq:BottCattaneo1}, \eqref{eq:BottCattaneo2};  note that the fiber is $S^4/\bZ_2$, instead of $S^4$.
So the anomaly polynomial $A_{E_8+\text{free}}(Q)$ of the E-string theory of rank $Q$ (plus free hyper multiplet ) is
\begin{align}
	A_{E_8+\text{free}}(Q)= \frac{Q^3}{6}p_2(N)+\frac{Q^2}{2}\chi_4(N)I_4+Q\left(\frac{1}{2}I_4^2-I_8\right).\label{eq:mainagain}
\end{align}  This gives \eqref{main}, once we rewrite $p_i(TX_{11})|_{X_6}$ in terms of $p_i(T)\equiv p_i(TX_6)$ and $p_i(N)$. 

We pause here to give two comments.  The first comment is 
on the behavior under $Q\to -Q$. When we change the orientation of the M5-branes, the preserved supersymmetry in 6d switches from \Nequals{(1,0)} to \Nequals{(0,1)}.  At the same time, this also exchanges the roles (the R-symmetry and the flavor symmetry) of two $\SU(2)$ groups in the decomposition $\SO(4)_N\simeq \SU(2)_L\times \SU(2)_R$. Correspondingly, the total anomaly should be multiplied by $-1$ when we make the transformation $Q\to -Q$ and $\chi_4(N)\to -\chi_4(N)$. This is indeed the case, as can be seen in \eqref{eq:mainagain}.

The second is on the `holographic' computation of anomalies.  We consider the spacetime of the form $\text{AdS}_7\times (S^4/\bZ_2)$, with the $E_8$ end-of-the-world brane at the boundary. The ansatz for the $G$ flux is $G=Qe_4+ e_0 I_4$; then the total Chern-Simons term on $\text{AdS}_7$, obtained by the integral over $S^4/\bZ_2$ of $S_{CGG}+S_{CI_8}$, is exactly the same as \eqref{intermediate} above. We chose not to use the holographic computation as the primary method, since that might have aroused the doubt whether it is exact in $Q$ or not.

\section{Checks}\label{checks}

\itemsep0pt
\topsep.5ex

\subsection{Terms that do not involve the normal bundle}
Let us perform some checks on the anomaly polynomial we obtained so far. As a first check, let us compute the terms that do not involve the normal bundle  using heterotic string theory on K3. 

Recall that the E-string theory of rank $Q$ is the low-energy theory of the small $E_8$ instanton of instanton number $Q$ in heterotic string theory.  To analyze this system, we consider a smooth K3 compactification where the two $E_8$ gauge bundles have instanton number $n_A$ and $n_B$ respectively, where $n_A+n_B=24$. This compactification was known as the $(n_A,n_B)$ heterotic compactification in the heyday of the second revolution. 

 We consider the choice of $(n_A,n_B)$ such that both gauge bundles are obtained by embedding smooth $\SU(2)$ gauge bundles. The unbroken gauge symmetry in 6d is then $E_{7,A}\times E_{7,B}$.
The low-energy matter content in 6d is 
i) 
 one supergravity multiplet and one tensor multiplet,
ii) 
twenty hypermultiplets describing the moduli of K3, 
iii) 
for both $E_{7,A}$ and $E_{7,B}$, we have \begin{itemize}
	\item an $E_7$ gauge multiplet,
	\item $n_{A,B}-4$  half-hypermultiplets in $\mathbf{56}$,
	\item $2n_{A,B}-3$ neutral hypermultiplets describing the moduli of $\SU(2)$ gauge bundle.
	\end{itemize}
The total anomaly polynomial of this matter content  is \begin{equation}
\frac1{16}(\tr R^2-\Tr  F_A^2 -\Tr  F_B^2)
(\tr R^2-(\frac{n_A}2-6)\Tr  F_A^2 -(\frac{n_B}2-6)\Tr  F_B^2)\label{hetanom}
\end{equation}  in our normalization, and is cancelled by the Green-Schwarz mechanism\footnote{See e.g.~\cite{Schwarz:1995zw,Berkooz:1996iz} for the details of the computation. Note that the normalization there is 16 times our normalization.}.

Now, let us collapse both the two smooth $\SU(2)$ instanton configurations into points. The unbroken gauge group is now $E_{8,A}\times E_{8,B}$, and for both $E_{8,A}$ and $E_{8,B}$, we have \begin{itemize}
\item an $E_8$ gauge multiplet 
\item and the E-string theory of rank $n_{A,B}$  as the ``matter content'',
\end{itemize}respectively. In addition, we have one supergravity multiplet, one tensor multiplet and twenty hypermultiplets as before.  The total anomaly should still be given by \eqref{hetanom}. From this we can compute the anomaly of the rank-$Q$ E-string theory, which turns out to be \begin{equation}
\frac{Q}{16}\left[
\frac1{12}\tr R^4 + \frac5{48}(\tr R^2)^2 - \frac12(\tr R^2)\Tr F^2 + \frac12(\Tr F^2)^2
\right]. 
\end{equation} Using $p_1=-\tr R^2/2$ and $p_2=(\tr R^2)^2/8-\tr R^4/4$, we find that this reproduces the terms in \eqref{main} independent of the normal bundle. 
 
\subsection{Coefficient of $(\Tr F^2)^2$}
In \cite{Witten:1996qz,Cheung:1997id}, the coefficient of $(\Tr F^2)^2$ in the anomaly polynomial was determined to be $Q$ times $-1/12$ of the contribution from an $E_8$ vector multiplet. 
Therefore, the term proportional to $(\Tr F^2)^2$ in the anomaly polynomial should be
 \begin{equation}
\frac{Q}{12}\frac{1}{4!}\tr_\text{adj} F^4=\frac{Q}{32}(\Tr F^2)^2,
\end{equation} agreeing with our central result \eqref{main}. 

Another manifestation of the same coefficient is as follows. In \cite{Sadov:1996zm}, the anomaly polynomial of an F-theory compactification on a Calabi-Yau 3-fold $X$ which is an elliptic fibration over the base $B$ was computed by dimensional reduction from 10 dimensions. The terms that only involve non-Abelian gauge fields can be stated easily: it is given in our normalization by \begin{equation}
I_8= \frac1{32}D\cdot D, \quad\text{where}\quad D=\sum_a D_a \Tr F^2_a \label{sadov}
\end{equation} where $D_a$ is the $a$-th divisor in $B$ supporting an F-theory 7-brane with a non-Abelian gauge symmetry,  $F_a$ is the corresponding curvature, and $D\cdot D$ is evaluated using the intersection pairing of the base $B$.

Now the E-string theory with $Q=1$ is obtained by blowing down a rational curve supporting no 7-brane singularity.  Suppose that the point after the blow-down is  an intersection of two divisors $D_{a}$, $D_b$ supporting 7-branes with gauge groups $G_a$, $G_b$ respectively. In \cite{Heckman:2013pva}, Appendix C, it was shown that the Lie algebra of $G_a\times G_b$ is always a subalgebra of the Lie algebra of $E_8$. 

This means that only the ``matter content'' charged under both $G_a$ and $G_b$ is the E-string theory itself. In view of \eqref{sadov}, this should contribute \begin{equation}
\frac1{16} \Tr F^2_a \Tr F^2_b
\end{equation} to the anomaly polynomial. This indeed arises from $(\Tr F^2)^2/32$ by replacing $\Tr F^2$ of $E_8$ by $\Tr F^2_a+\Tr F^2_b$ of $G_a\times G_b$.

\subsection{Behavior when $Q=1$}
As a final check, consider the behavior of our anomaly polynomial \eqref{main} when $Q=1$. 
For general $Q$, the theory definitely has the flavor symmetry $E_8\times \SU(2)_L$.
In particular, this $\SU(2)_L$ acts on the free hypermultiplet describing the center-of-mass motion of $Q$ M5-branes within the $E_8$ end-of-the-world brane: the hypermultiplet is a half-hypermultiplet in the doublet of $\SU(2)_L$.

For $Q=1$,  however, it is believed that the $\SU(2)_L$ does not act on the interacting part of the E-string theory.  Such an additional $\SU(2)_L$ flavor symmetry was never seen in the analysis of the $T^2$ compactification of the rank-1 E-string theory.
Note also that the Higgs branch of the rank-$Q$ E-string theory is the moduli space of $Q$ instantons of $E_8$.  When $Q=1$, the Higgs branch of the interacting part of the E-string theory is the centered moduli space of 1-instanton of $E_8$. The $\SU(2)_R$-symmetry acts nontrivially by rotating three complex structures, whereas $\SU(2)_L$ does not act on the centered 1-instanton moduli space. 
Due to the same fact, the $\SU(2)_L$ symmetry on 4d  \Nequals{2} superconformal theories on $Q$ D3-branes probing a 7-brane is nontrivial only when $Q>1$ \cite{Aharony:2007dj}.

Therefore, we expect that the anomaly polynomial of the E-string theory with rank 1, without the free hypermultiplet, should be completely free  of the characteristic class $c_2(L)$ of $\SU(2)_L$. Indeed, subtracting the contribution of the free hypermultiplet \eqref{freehyper} from the total anomaly \eqref{main}, and setting $Q=1$, we see that all the coefficients of monomials involving $c_2(L)$ vanish. There are four such potentially non-vanishing terms, namely $c_2(L)^2$, $c_2(L)c_2(R)$, $c_2(L)p_1(T)$ and $c_2(L)\Tr F^2$, making the check  rather nontrivial.

\section{Conclusions and future directions}\label{conclusions}
In this paper, we computed the anomaly polynomial of the E-string theory of general rank $Q$, by combining the methods of Ho\v rava and Witten and of Freed, Harvey, Minasian and Moore. The result was given in \eqref{main}, and contained terms of order $Q^3$, $Q^2$ and $Q$.  We performed a few consistency checks of our result, by comparing it against known properties of the E-string theory available in the literature. 

As a generalization of our work, it would be worth while to compute the anomaly polynomials of other \Nequals{(1,0)} superconformal theories.
One general way to have \Nequals{(1,0)} theories is to consider  M5-branes or  heterotic $E_8$ small instantons on an ALE singularity; similarly, we can put heterotic $\SO(32)$ instantons on an ALE singularity. These theories  were studied  using various duality frames.  Namely, F-theory was used in \cite{Aspinwall:1996vc,Aspinwall:1997ye}, NS5-branes on ALE  in    \cite{Intriligator:1997kq,Blum:1997fw,Blum:1997mm}, and  D8-D6-NS5 brane setup   in  \cite{Brunner:1997gf,Hanany:1997gh}.  
The ones which have realizations with M5-branes are rather similar to the E-string theories of general rank treated in this paper, and it would not be very hard to compute their anomaly polynomials.  

Holographic duals of some of these theories were also constructed using massive type IIA supergravity in \cite{Gaiotto:2014lca}. It would be interesting to compute their anomaly polynomials independently and compare with the results from holography.  

A classification scheme of 6d \Nequals{(1,0)} superconformal theories was given in \cite{Heckman:2013pva} using F-theory.  It would be nice to obtain a general formula for the anomaly polynomials in terms of the F-theory data. The `normal bundle' part of the anomaly is, however, rather hard to see in the F-theoretic approach, and the authors do not know how to proceed at present. 

It would also be interesting to study the anomaly inflow to the worldsheet of the self-dual string from the bulk of the E-string theory; note that such inflow to the string worldsheet in the case of \Nequals{(2,0)} theories was studied in  \cite{Henningson:2004dh,Berman:2004ew}.

Another direction of research is to study the compactification of the E-string theory of general rank on a Riemann surface, possibly with punctures.  This should give rise to a large class of 4d \Nequals{1} theories with $E_8$ flavor symmetry, and should be an \Nequals{1} analogue of Gaiotto's construction \cite{Gaiotto:2009we}.  With the anomaly polynomials obtained in this paper, we can at least compute the central charges $a$ and $c$ of these theories, assuming that there is no emergent $\U(1)$ mixing with the superconformal R-symmetry.

Of course, it would be more interesting if we could systematically understand compactifications of other 6d  \Nequals{(1,0)} theories discussed above on general Riemann surfaces and on higher-dimensional manifolds, too.  There seems to be many interesting structures ready to be uncovered in the near future.

\section*{Acknowledgements}
KO and HS are partially supported by the Programs for Leading Graduate Schools, MEXT, Japan,
via the Advanced Leading Graduate Course for Photon Science
and via the  Leading Graduate Course for Frontiers of Mathematical Sciences and Physics, respectively. 
KO is also supported by JSPS Research Fellowship for Young Scientists.
YT is  supported in part by JSPS Grant-in-Aid for Scientific Research No. 25870159,
and in part by WPI Initiative, MEXT, Japan at IPMU, the University of Tokyo.

\appendix

\section{Global angular forms}\label{appendix}
In this appendix, we recall the properties of a smoothed-out version of  differential forms with delta-function support. They are constructed using the help of the so-called global angular forms. 

Let $M$ denote an oriented manifold and $E$ be an oriented rank $2k+1$ real vector bundle over $M$. Assume that $E$ admits  a metric and a connection $\Theta$ compatible with its metric.
Denote its zero section by $s_0$.
We can construct an $S^{2k}$ bundle $\pi:S(E)\to M$ which is homeomorphic to $E_0 = E\setminus s_0(M)$
by assigning each point $p$ of $M$ a unit sphere in the fibre of $E$ around $s_0(p)$.

Then, we can construct a form $e_{2k}$ on $S(E)$ which satisfies following properties:
\begin{itemize}
	\item $e_{2k}$ is a globally well-defined $2k$-form on $S(E)$.
	\item $ d e_{2k} = 0$.
	\item $\int_{\pi^{-1}(p)}e_{2k}|_{\pi^{-1}(p)}= 2$ for any point $p$ of $M$. In other words, $\pi_* e_{2k} =2$.
\end{itemize}
Let $\rho$ be a compactly supported function on $E$ which satisfies $\rho(s(p))=-1$.
We can explicitly write a Thom class $\Phi(E)$ of the bundle $E$, the smooth analogue of $\delta (M\hookrightarrow E)$, as
\begin{align}
	\Phi(E)= d \rho\:e_{2k}/2.
\end{align}
Here identify the form $e_{2k}$ on $S(E)$ and its pullback in terms of the homeomorphism $S(E)\simeq E_0$.

We can apply the usual decent notation:
\begin{align}
	d e_{2k-1}^{(0)}=e_{2k},\quad \delta e_{2k-1}^{(0)}= e_{2k-2}^{(1)}.
\end{align}
Here $\delta$ denote a $SO(2k+1)$ gauge transformation associated with the connection $\Theta$.

Let us concentrate on the cases $k=0$ and $k=2$, which are relevant for our calculation. $e_0$ is just a step function whose value is $+1$ or $-1$. The explicit form of $e_{4}$ is given by
\begin{eqnarray}
e_{4}=\frac{1}{32\pi^{2}}\epsilon_{a_{1}\cdots a_{5}}&\biggl{[}&(D\hat{y})^{a_{1}}(D\hat{y})^{a_{2}}(D\hat{y})^{a_{3}}(D\hat{y})^{a_{4}}\hat{y}^{a_{5}} \nonumber \\
 &-&2F^{a_{1}a_{2}}(D\hat{y}^{a_{3}})(D\hat{y})^{a_{4}}\hat{y}^{a_{5}} + F^{a_{1}a_{2}}F^{a_{3}a_{4}}\hat{y}^{a_{5}}\biggr{]}
\end{eqnarray}
Here, $a_{i}=1\cdots 5$ labels the fiber coordinates and $\hat y^{a_{i}}$ are coordinates of the unit sphere $S^4$. A covariant derivate and 2-form is defined using the connection $\Theta$ by
\begin{align}
 D\hat{y}^{a}&=d\hat{y}^{a}-\Theta^{ab}\hat{y}^{b} , &
 F^{ab}&=d\Theta^{ab} - \Theta^{ac}\wedge \Theta^{cb}.
\end{align} 
Using this explicit form, we can prove the formulae \begin{equation}
\pi_{*}(e_{4})=2,\quad \pi_{*}(e_{4}^{3})=2p_{2}(E).\label{eq:BottCattaneo1}
\end{equation}
The formula $\pi_{*}(e^3_4)=2p_2(N)$ is first proved by Bott and Cattaneo \cite{Bott:1997in}.

When the $\SO(5)$ connection reduces to $\SO(4)$, we can consider $e_0$ and $e_4$ at the same time; $e_0$ is a step function taking $+1$  and $-1$ on the northern and the southern hemispheres of $S^4$, respectively. Then we have 
\begin{equation}
 \pi_{*}(e_{4}e_{0}^2)=2,\quad \pi_{*}(e_{4}^{2}e_{0})=2\chi_{4}(F),\label{eq:BottCattaneo2}
\end{equation}
where $\chi_4(F)$ is the Euler class of the rank 4 bundle $F$ which is associated with the $\SO(4)$ connection.

\bibliographystyle{ytphys}
\bibliography{ref}

\end{document}